\documentclass{article}

\usepackage{color}
\usepackage{soul}

\usepackage{subfigure}
\usepackage{hyperref}
\usepackage{graphicx}
\usepackage{amssymb}
\usepackage{amsmath}
\usepackage{times}

\usepackage{amsthm}

\usepackage{fullpage}

\usepackage{xspace}
\usepackage{algorithm}
\usepackage[noend]{algpseudocode}

\newcommand{\ceil}[1]{\left\lceil #1 \right\rceil}
\newcommand{\floor}[1]{\left\lfloor #1 \right\rfloor}

\begin{document}
\title{Parallel Space-Time Kernel Density Estimation}

\author{Erik  Saule$^\dag$, Dinesh Panchananam$^\dag$, Alexander Hohl$^\ddag$, Wenwu Tang$^\ddag$, Eric Delmelle$^\ddag$\\
  $^\dag$ Dept. of Computer Science, $^\ddag$Dept. of Geography and Earth Sciences\\
UNC Charlotte\\
Email: {\it \{esaule,dpanchan,ahohl,wtang4,eric.delmelle\}@uncc.edu}
}

\maketitle

\begin{abstract}
  The exponential growth of available data has
  increased the need for interactive exploratory analysis. Dataset
  can no longer be understood through  manual crawling  and
 simple statistics. In Geographical Information Systems
  (GIS), the dataset is often composed of events localized in space
  and time; and visualizing such a dataset involves building a map of
  where the events occurred.

  We focus in this paper on events that are localized among three
  dimensions (latitude, longitude, and time), and on computing the
  first step of the visualization pipeline, space-time kernel density
  estimation (STKDE), which is most computationally expensive.  Starting from
  a gold standard implementation, we show how algorithm design and
  engineering, parallel decomposition, and scheduling can be applied
  to bring near real-time computing to space-time kernel density
  estimation. We validate our techniques on real world datasets
  extracted from infectious disease, social media, and ornithology.
\end{abstract}

\maketitle

\section{Introduction}

The rapid propagation of infectious diseases (e.g. zika, Ebola, H1N1,
dengue fever) is conducive to serious, epidemic outbreaks, posing a
threat to vulnerable populations. Such diseases have complex
transmission cycles, and effective public health responses require the
ability to monitor outbreaks in a timely
manner~\cite{eisen2011using}. Space-time statistics facilitate the
discovery of disease dynamics including rate of spread, seasonal
cyclic patterns, direction, intensity (i.e. clusters), and risk of
diffusion to new regions.  However, obtaining accurate results from
space-time statistics is computationally very demanding, which is
problematic when public health interventions are promptly needed. The
issues of computational efforts are exacerbated with spatiotemporal
datasets of increasing size, diversity and
availability~\cite{grubesic2014spatial}. High-performance computing
reduces the effort required to identify these patterns, however
heterogeneity in the data must be accounted for~\cite{Hohl16}. \looseness=-1

Epidemiology is only one of the application domains where it is
important to understand how events occurring at different locations and
different times form clusters. Political analysis, social media
analysis, or the study of animal migration also require the
understanding of spatial and temporal locality of events. The massive
amount of data we see nowadays is often analyzed using complex
models. But before these can be constructed, data scientists  need to
interactively visualize and explore the data to understand its
structure. \looseness=-1

In this paper, we present the space-time kernel density estimation
application which essentially builds a 3D density map of events
located in space and time. This problem is computationally expensive
using existing algorithms. That is why we developed better sequential
algorithms for this problem that reduced the complexity by orders of
magnitude. And to bring the runtime in the realm of near real-time, we
designed parallel strategies for shared memory machines.

Section~\ref{sec:stkde} presents the STKDE application along with a
reference implementation that uses a voxel-based algorithm VB to
compute STKDE. In Section~\ref{sec:algo-seq}, we investigate the
sequential problem and develop a point-based algorithm PB that
significantly reduce the complexity compared to VB. We also identify
invariants linked to the structure of the kernel density estimate
function that allows to reduce the computational cost by an additional
factor and build the PB-SYM algorithm based on that observation. We
develop two parallelization strategies (PB-SYM-DR and PB-SYM-DD) in
Section~\ref{sec:pb-sym-domain} that are designed to make the
computation pleasingly parallel. However both strategies are not work-efficient
and can cause significant work overhead in some cases. That
is why we introduce PB-SYM-PD in Section~\ref{sec:pb-sym-point} that
provide a work-efficient decomposition of the points. However that
algorithm has internal dependencies which can induce a long critical path
and cause load imbalance; and we develop PB-SYM-PD-SCHED to leverage
alternative ordering of the work to reduce the critical path. To
reduce the critical path further, we propose the PB-SYM-PD-REP algorithm which
introduces some work
overhead only where is needed to achieve a short critical path.
Section~\ref{sec:experiments} presents experimental
results on 4 different datasets and 21 instances extracted from
these datasets. The experiments highlights that the sequential
strategies are efficient. The parallel strategies have their pros and
cons and different strategies appear to be best in different
scenarios. Section~\ref{sec:soa} discusses the applicability of related problems and
techniques.

\section{Space-Time Kernel Density Estimation}
\label{sec:stkde}

\subsection{Description}

Space-time kernel density (STKDE) is used for identifying
spatiotemporal patterns in datasets. It is a temporal
extension of the traditional 2D kernel density
estimation~\cite{Silverman86} which generates density surface
(``heatmap'') from a set of $n$ points located in a geographic space. The resulting density
estimates are visualized within the space-time cube framework using two spatial $(x,
y)$ and a temporal dimension $(t)$~\cite{Nakaya10}. STKDE creates a discretized 3D
 volume where each voxel (3D equivalent of a pixel) is assigned a
density estimate based on the surrounding points. The space-time
density is estimated using (following the notations
of~\cite{Hohl16}):

$$ \hat{f} (x, y, t) = \frac{1}{n h_s^2 h_t} \sum_{i | d_i < h_s, t_i<h_t}
k_s (\frac{x-x_i}{h_s},\frac{y-y_i}{h_s}) k_t(\frac{t-t_i}{h_t})$$

Density $\hat{f} (x, y, t)$ of each voxel is determined by number and distance of events (points)
$(x_i, y_i, t_i)$ within its vicinity, which is conceptualized by a
cylinder. The spatial bandwidth $h_s$ forms a circle which, due to the
orthogonal relationship between space and time, is extended to a
cylinder by temporal bandwidth $h_t$. For every event inside the
cylinder, the spatial ($d_i$) and temporal ($t_i$) distances are smaller
than $h_s$ and $h_t$. Therefore, the event receives a weight based on the
kernel functions $k_s$ and $k_t$ (distance decay):

$$k_s (u,v) = \frac{\pi}{2} (1-u)^2 (1-v)^2$$
$$k_t (w) = \frac{3}{4} (1-w)^2$$

\begin{figure}
  \subfigure[$h_s = 2500m$, $h_t = 14 days$]{\includegraphics[width=.49\linewidth]{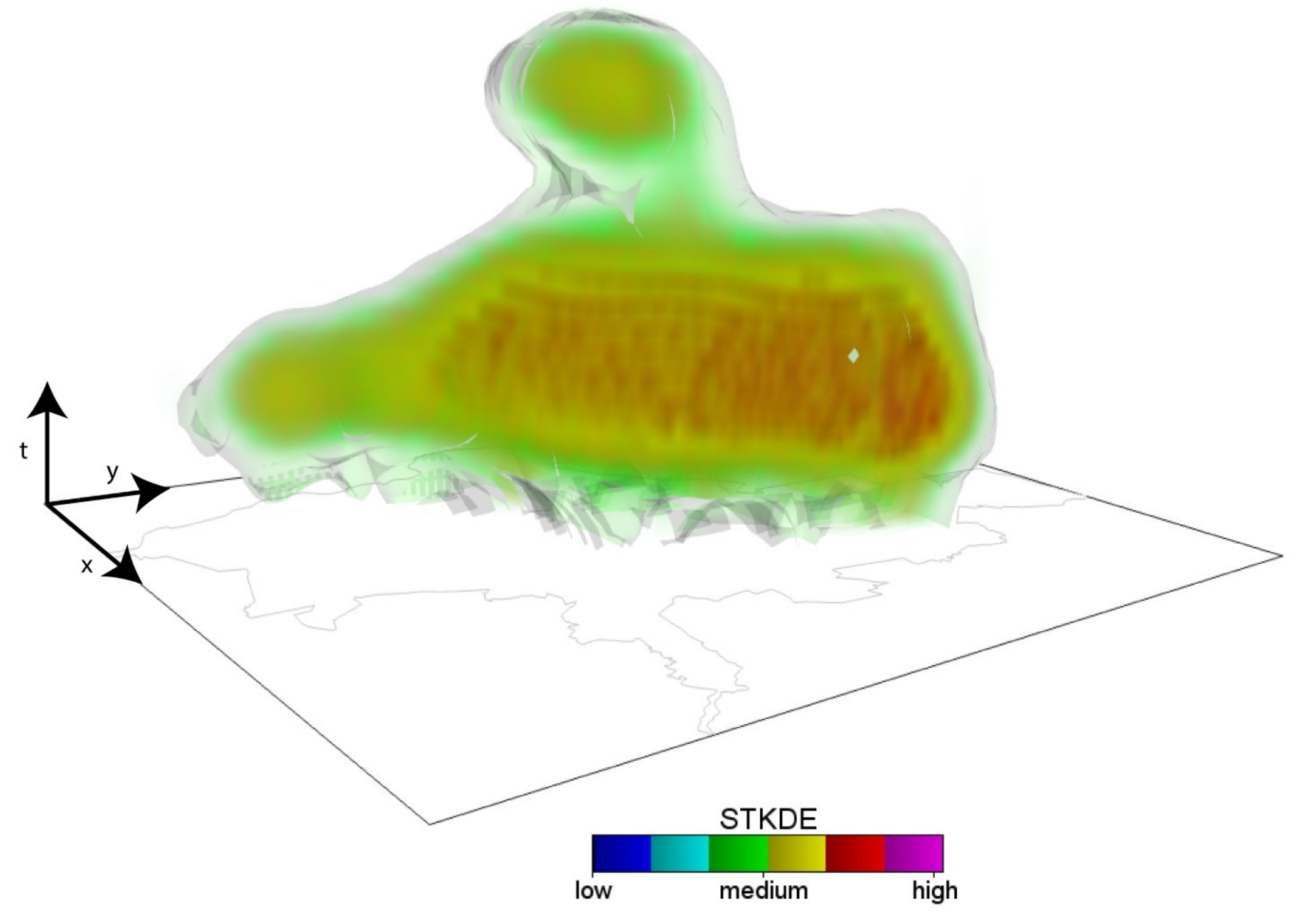}}
%
\subfigure[$h_s=500m$, $h_t = 7 days$]{\includegraphics[width=.49\linewidth]{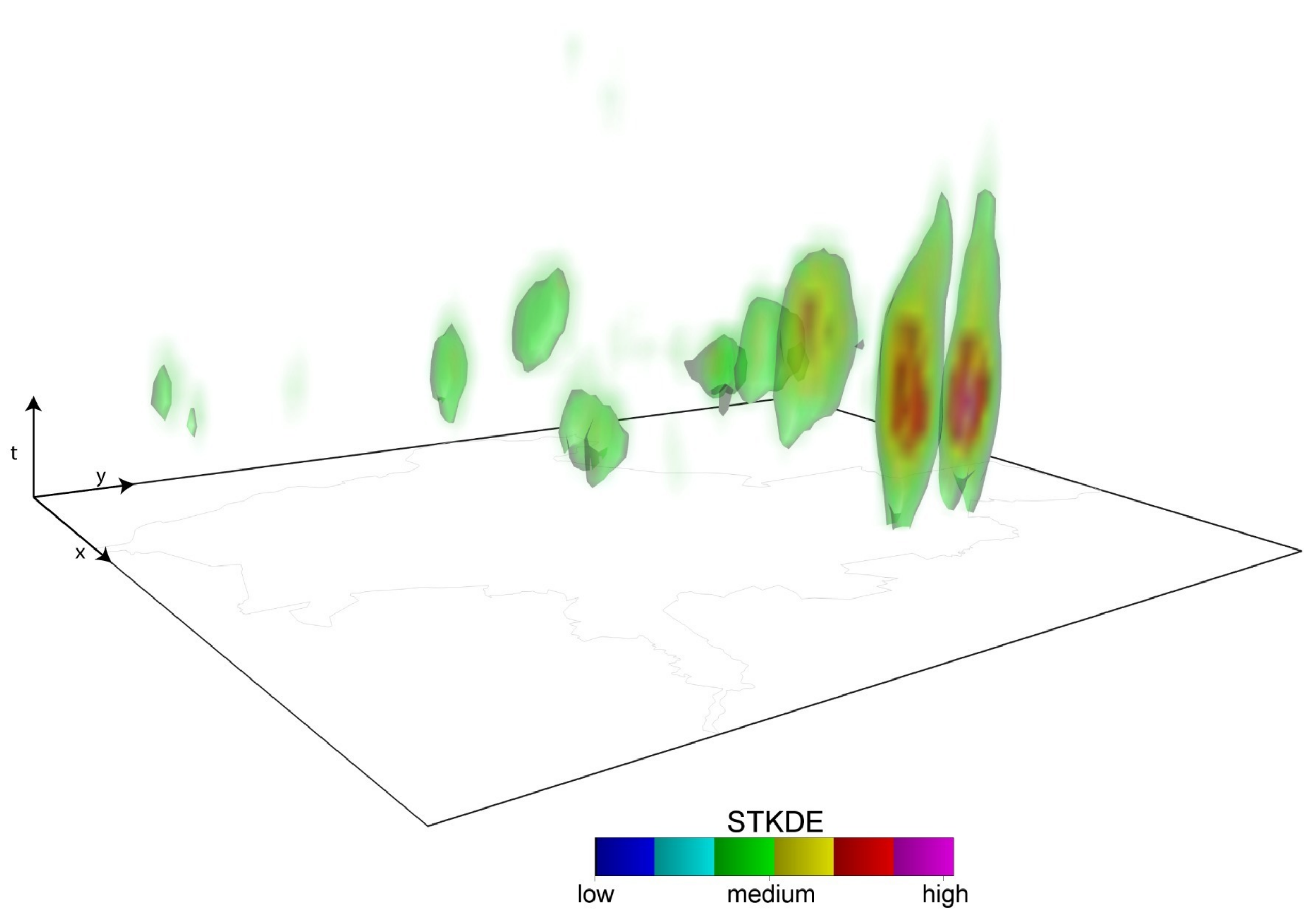}}
  \caption{Visualization of Dengue fever cases in Cali, Colombia in
    2010 and 2011 for different spatial bandwidth and temporal bandwidth.}
  \label{fig:stkde-ex}
\end{figure}

\begin{figure}
  \centering
  \includegraphics[width=.5\linewidth]{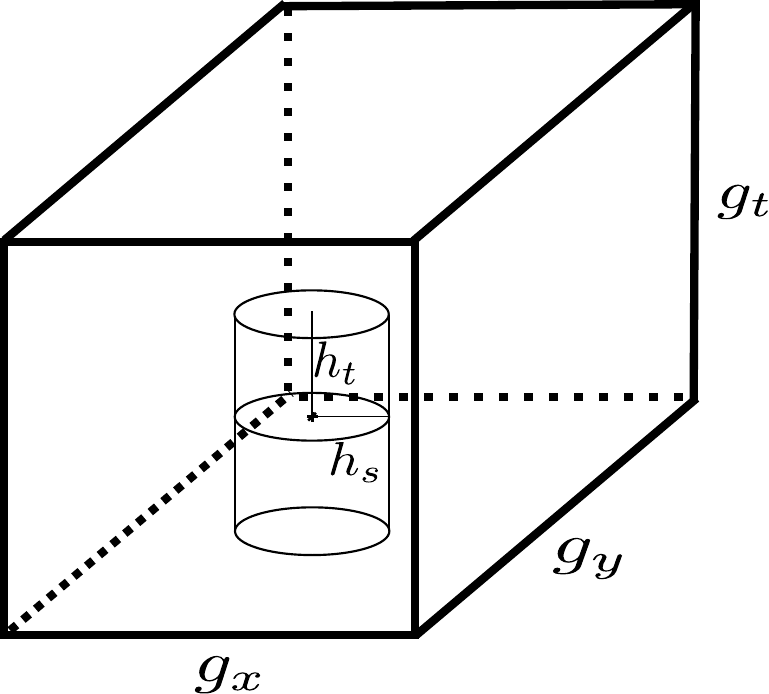}
  \caption{The computation of STKDE happens in a domain space of size
    $g_x,g_y,g_t$. Each point impacts the neighboring space at a
    distance $h_s$ in space and $h_t$ in time; forming a cylinder of
    diameter $2 h_s$ and of height $2 h_t$.}
  \label{fig:stkde}
\end{figure}

Figure~\ref{fig:stkde-ex} illustrates how varying the bandwidths used
in STKDE helps focusing the graphical visualization of Dengue fever
cases in Cali, Colombia~\cite{Delmelle14}. Figure~\ref{fig:stkde} shows the impact
of a single point on the neighboring space.

Computationally, the domain of size $g_x$, $g_y$, $g_t$ is discretized
in voxels using a spatial resolution $sres$ and a temporal resolution
$tres$. Therefore, the domain is represented by a grid of size
$G_x=\ceil{\frac{g_x}{sres}}$, $G_y=\ceil{\frac{g_y}{sres}}$,
$G_t=\ceil{\frac{g_t}{tres}}$. Each point causes a density increase in
the voxels that are within a cylinder centered on the point, of
radius equal to the spatial bandwidth in voxels
$H_s=\ceil{\frac{h_s}{sres}}$, and of half height equal to the
temporal bandwidth $H_t=\ceil{\frac{h_t}{tres}}$.

All the notations are summarized in Table~\ref{tab:notation}. As a
convention all uppercase notations denote quantities in voxels and all
lowercase notations denote quantities in domain space.

\begin{table}
  \centering
  \begin{tabular}{|l|l|}
    \hline 
    $n$ & Number of points\\
    $s = (x,y,t)$ & A voxel and sampling coordinate\\
    $(x_i, y_i, t_i)$ & Coordinate of point $i$\\
    $h_s$, $h_t$ & Spatial and temporal bandwidth \\ 
    $g_x, g_y, g_t$ & Real size of the domain (in meters) \\
    \hline
    $sres$, $tres$ & Resolution (in meters) \\
    $s = (X,Y,T)$ & A voxel in voxel space\\
    $(X_i, Y_i, T_i)$ & Voxel of point $i$\\
    $G_x, G_y, G_t$ & Size of the domain (in voxels)\\
    $H_s, H_t$ & Bandwidth (in voxels) \\
    \hline 
  \end{tabular}
  \caption{Notations}
  \label{tab:notation}
\end{table}

\subsection{Gold Standard Implementation}

The gold standard implementation (for instance from~\cite{Hohl16}) follows the exact definition of STKDE
as it is given above. It is a voxel-based algorithm we call VB. For
each voxel, VB finds the points within the temporal and spatial bandwidths and calculates the contribution
to the density of this voxel. The pseudo code is given in Algorithm~\ref{alg:vb}.

\begin{algorithm}
\caption{\textsc{VB}}
\begin{algorithmic}
\For{all voxels $s=(x,y,t)$}
\State $sum = 0$
  \For{all points $i$ at $x_i, y_i, t_i$}
    \If{$\sqrt{(x_i-x)^2+(y_i-y)^2}< h_s$ and $|t_i-t| \leq h_t$}
    \State $sum += k_s(\frac{x-x_i}{h_s},\frac{y-y_i}{h_s}) k_t(\frac{t-t_i}{h_t})$
    \EndIf
  \EndFor
  \State $stkde[X][Y][T] = \frac{sum}{n h_s^2h_t}$
\EndFor
\end{algorithmic}
\label{alg:vb}
\end{algorithm}

The algorithm performs $\theta(G_x G_y G_t n)$ distance tests and
computes $\theta(n H_s^2 H_t )$ densities. Since $H_s$ is smaller than
$G_x$ and $G_y$ and since $H_t$ is smaller than $G_t$, the complexity of the
algorithm is $\theta(G_x G_y G_t n )$ and it requires $\theta(G_x G_y G_t)$
memory.

\section{Algorithm Design and Engineering}

\label{sec:algo-seq}

\subsection{Point-based Algorithm}
The voxel-based algorithm suffers from a massive cost of computing
distances. The gold implementation reduces the number of distance computation by 
decomposing the domain, but that still incurs many distance calculations.
Most of these calculations are unnecessary since we know
where each point will radiate density. The point-based
algorithm PB, given in Algorithm~\ref{alg:pb}, leverages that
property.

\begin{algorithm}
\caption{\textsc{PB}}
\begin{algorithmic}
\For{all voxels $s=(x,y,t)$}
  \State $stkde[X][Y][T] = 0$
\EndFor
\For{each points $i$ at $x_i, y_i, t_i$}
  \For{$X_i - H_s\leq X \leq X_i+H_s$}
    \For{$Y_i - H_s\leq Y \leq Y_i+H_s$}
      \For{$T_i - T_s\leq T \leq T_i+H_s$}
        \If{$\sqrt{(x_i-x)^2+(y_i-y)^2}< h_s$ and $|t_i-t| \leq h_t$}
        \State $stkde[X][Y][T] += \frac{k_s(\frac{x-x_i}{h_s},\frac{y-y_i}{h_s}) k_t(\frac{t-t_i}{h_t})}{n h_s^2h_t}$
        \EndIf
      \EndFor
    \EndFor
  \EndFor
\EndFor
\end{algorithmic}
\label{alg:pb}
\end{algorithm}

The algorithm PB incurs mostly two costs. Initializing the memory which costs
$\Theta(G_x G_y G_t)$ and computing the impact of each point
$\Theta(n H_s^2 H_t)$. Note that it is possible that either of these
two terms is significantly larger than the other. Therefore the
complexity of the algorithm is $\Theta(G_x G_y G_t + n H_s^2 H_t)$.

\subsection{Exploiting Symmetries}

When the bandwidth is large, computing the density contribution of each
point each for voxel in the bandwidth is the most expensive
operation. Each of these $n (2  H_s)^2  2 H_t$ calculations costs
approximately 40 floating point operations. (The exact cost is
difficult to estimate since there are floating point additions,
multiplications, divisions, and square root operations).

\begin{figure}
  \centering
  \includegraphics[width=.6\linewidth]{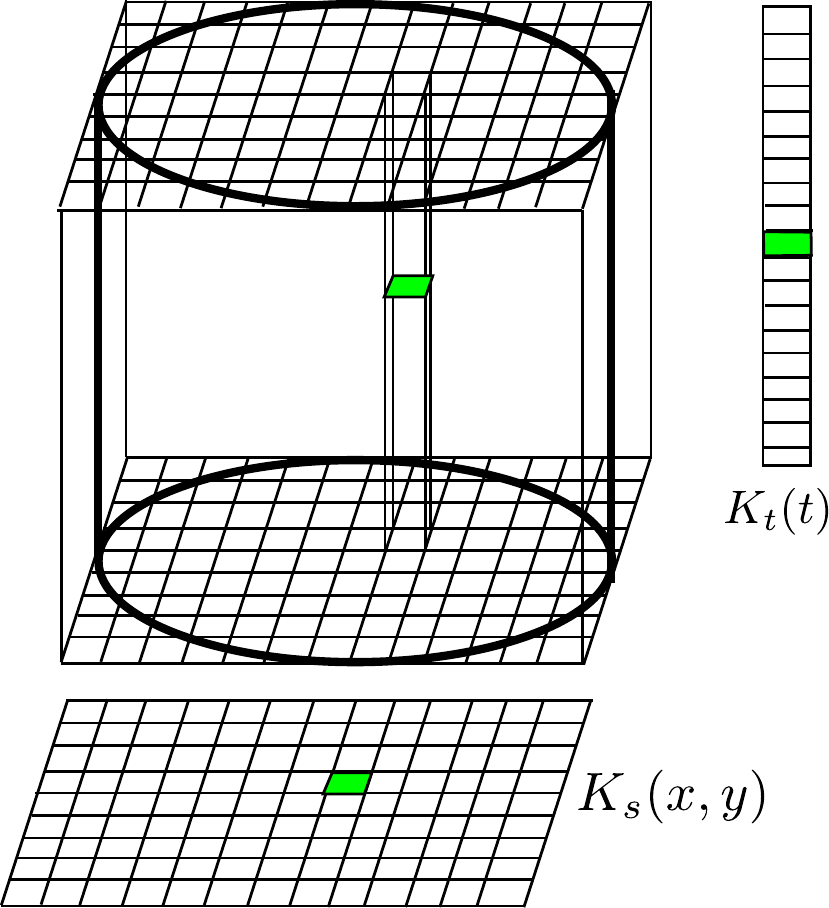}
  \caption{The contribution to the density estimate of each voxel of a
    cylinder can be decomposed in two terms, one temporally invariant
    $K_s(x,y)$ and one spatially invariant $K_t(t)$.}
  \label{fig:pb-sym-decomposition}
\end{figure}

One can remark that the calculation is mostly redundant since for one
particular point, its contribution to the neighboring voxels can be
decomposed in two components. $K_s(x,y)$  only depends on the
spatial coordinate of the voxel with respect to the point and does not
depend on its temporal coordinate. And $K_t(t)$  only depends on
the temporal coordinate and not on the spatial coordinates. See
Figure~\ref{fig:pb-sym-decomposition} for a graphical depiction.

Therefore, we write an algorithm to leverage these symmetries in
the problem. We present three variants, the PB-DISK variant ensures that
the invariant on the spatial domain is only computed once. The PB-BAR
variant ensures that the invariant on the temporal domain is only
computed once. The PB-SYM variant computes independently the spatial
invariant and the temporal invariant and then adds the density caused
by the point by taking the product of the two invariants. (See pseudocode in Algorithm~\ref{alg:pb-sym}.) 
Notice that leveraging the symmetries in the calculation does not change the
complexity (PB-SYM still has a complexity of $\Theta(G_x 
G_y  G_t + n  H_s^2  H_t)$) but reduces the number of flops.

\begin{algorithm}
\caption{\textsc{PB-SYM}}
\begin{algorithmic}
\For{all voxels $s=(x,y,t)$}
  \State $stkde[X][Y][T] = 0$
\EndFor
\For{each points $i$ at $x_i, y_i, t_i$}
  \For{$X_i - H_s\leq X \leq X_i+H_s$}
    \For{$Y_i - H_s\leq Y \leq Y_i+H_s$}
      \If{$\sqrt{(x_i-x)^2+(y_i-y)^2}< h_s$ }
        \State $K_s[X][Y] = \frac{k_s(\frac{x-x_i}{h_s},\frac{y-y_i}{h_s})}{n h_s^2h_t}$
      \Else
        \State $K_s[X][Y] = 0$
      \EndIf
    \EndFor
  \EndFor
  \For{$T_i - T_s\leq T \leq T_i+H_s$}
    \If{$|t_i-t| \leq h_t$}
      \State $K_t[T] = k_t(\frac{t-t_i}{h_t})$
    \Else
      \State $K_t[T] = 0$
    \EndIf
  \EndFor
  \For{$X_i - H_s\leq X \leq X_i+H_s$}
    \For{$Y_i - H_s\leq Y \leq Y_i+H_s$}
      \For{$T_i - T_s\leq T \leq T_i+H_s$}
        \State $stkde[X][Y][T] +=  K_s[X][Y] K_t[T]$
      \EndFor
    \EndFor
  \EndFor
\EndFor
\end{algorithmic}
\label{alg:pb-sym}
\end{algorithm}

Notice that this exploitation of symmetries is not possible in a
voxel-based algorithm.

\label{sec:algo-pb-sym}

\section{Domain-based parallelism}
\label{sec:pb-sym-domain}

\subsection{Domain Replication}

The simplest way to parallelize this kind of computation is to split
the points equally among the $P$ different computational
units. Though, if two close-by points are computed
simultaneously, the cylinders of density around them might intersect and
there is a race condition.

The PB-SYM-DR algorithm solves the data race issue by having each of
the $P$ computational unit aggregate the result on a local copy of the
domain. Once all the points have been processed, the $P$ copies need
to be summed. The pseudo code of PB-SYM-DR is given in
Algorithm~\ref{alg:pb-sym-dr}.

This algorithm has a memory requirement of $\Theta (P  G_x  G_y  G_t)$
and a parallel amount of work of $\Theta (P  G_x  G_y  G_t + n  H_s^2
 H_t)$. Fortunately, the increase in work enables the computation to be
pleasingly parallel. (There are actually in three pleasingly parallel phases:
initializing the memory, processing the points, and reducing the partial results.)

\begin{algorithm}
\caption{\textsc{PB-SYM-DR}}
\begin{algorithmic}
\For{all processor $p \leq P$ in parallel}
  \For{all voxels $s=(x,y,t)$}
    \State $stkdel[p][X][Y][T] = 0$
  \EndFor
\EndFor
\For{each points $i$ at $x_i, y_i, t_i$ distributed among the $P$ processors}
  \State Processor $p$ processes $i$ using PB-SYM and aggregate it in $stkdel[p]$
\EndFor
\For{all voxels $s=(x,y,t)$ in parallel}
  \State $stkde[X][Y][T] = 0$
  \For{all processor $p \leq P$}
    \State $stkde[X][Y][T] += stkdel[p][X][Y][T]$
  \EndFor
\EndFor
\end{algorithmic}
\label{alg:pb-sym-dr}
\end{algorithm}

\subsection{Domain Decomposition}
\label{sec:pb-sym-dd}

Another strategy to avoid the data race is to compute 
voxel intensity  independently for different subdomains. We call this method PB-SYM-DD (Algorithm~\ref{alg:pb-sym-dd}).
It splits the domain in AxBxC subdomains and each subdomain is
associated the points which cylinder intersects with a voxel of
the subdomain. Then the algorithm proceeds with handling each
subdomain independently by only considering the impact of the
data-point on voxels within the subdomain.

\begin{algorithm}
\caption{\textsc{PB-SYM-DD}}
\begin{algorithmic}
\For{all points $i=(x_i,y_i,t_i)$}
  \For{each subdomain $(a,b,c)$}
        \If{$(X,Y,T) \pm (H_s,H_s,H_T)$ intersects \\$(\floor{\frac{a G_x}{A}}:\ceil{\frac{(a+1) G_x}{A}}, \floor{\frac{b G_y}{B}}:\ceil{\frac{(b+1) G_y}{B}}, \floor{\frac{c G_t}{C}}:\ceil{\frac{(c+1) G_t}{C}})$}
          \State $localpoints[a][b][c].add(x_i,y_i,t_i)$
        \EndIf
  \EndFor
\EndFor
\For{each subdomain $(a,b,c)$ in parallel}
  \State Process subdomain $(a,b,c)$ using PB-SYM
\EndFor
\end{algorithmic}
\label{alg:pb-sym-dd}
\end{algorithm}

\begin{figure}
  \centering
  \includegraphics[width=.6\linewidth]{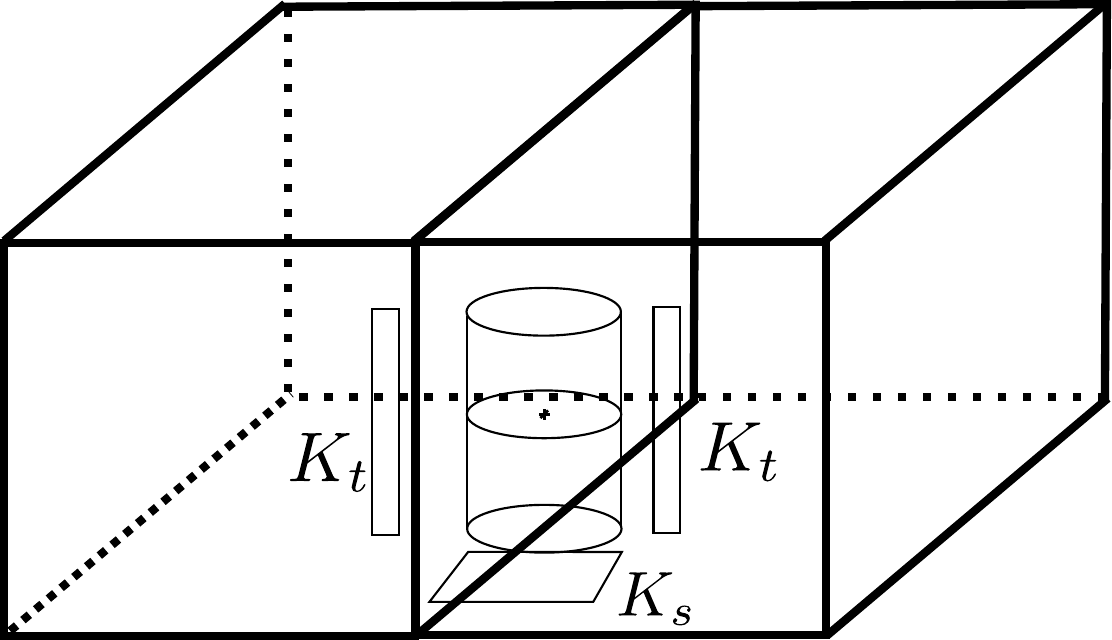}
  \caption{When a cylinder is split among two subdomains, PB-SYM-DD
    causes an overhead because one of the invariant needs to be
    recomputed. Here each subdomains compute part of the temporal
    invariant $K_s$, but both need to compute the spatial invariant $K_t$.}
  \label{fig:pb-sym-dd-overhead}
\end{figure}

The main problem with Domain Decomposition is that it requires some
points to be replicated across multiple subdomain which incurs
additional work. Indeed, if a point is replicated in multiple subdomains,
its cylinder is split across multiple subdomains as well. Assuming the split is
temporal, then using PB-SYM in each subdomain requires computing
part of the temporal invariant, the entire spatial invariant, and part
of the cylinder. Therefore both subdomains need to compute the entire
spatial invariant. See Figure~\ref{fig:pb-sym-dd-overhead} for a
graphical depiction of this phenomenon.

If the algorithms keeps the number of subdomains small such that each
point is replicated less than a constant number of times, then
the work expressed in Big-Oh notation does not change.
(This can be achieved by having subdomains larger than
$H_s \times H_s \times H_t$.) However, the practical amount of work can
increase by a non-negligable factor. 

Another issue with Domain Decomposition is load imbalance. The
points are unlikely to be equally distributed in the domain space, but
more likely clustered around some locations. The subdomain containing
a cluster will have a significantly higher work than other
subdomains. And this work is not executed in parallel. One could
decompose the domain more, but replicated point 
will incur an even higher overhead which may end up nullifying
the gain in load balance.\looseness=-1

\section{Point-based parallelism}
\label{sec:pb-sym-point}

\subsection{Point Decomposition}
\label{sec:pb-sym-pd}

Both of the previous strategies increase the amount of work
significantly. We want to achieve parallelism without cutting a
cylinder and without requiring each thread to have its own memory
space. To achieve that goal, we propose the PB-SYM-PD algorithm that
decomposes the points in sets that can be safely performed
simultaneously.

\begin{figure}
  \centering
  \includegraphics[width=.5\linewidth]{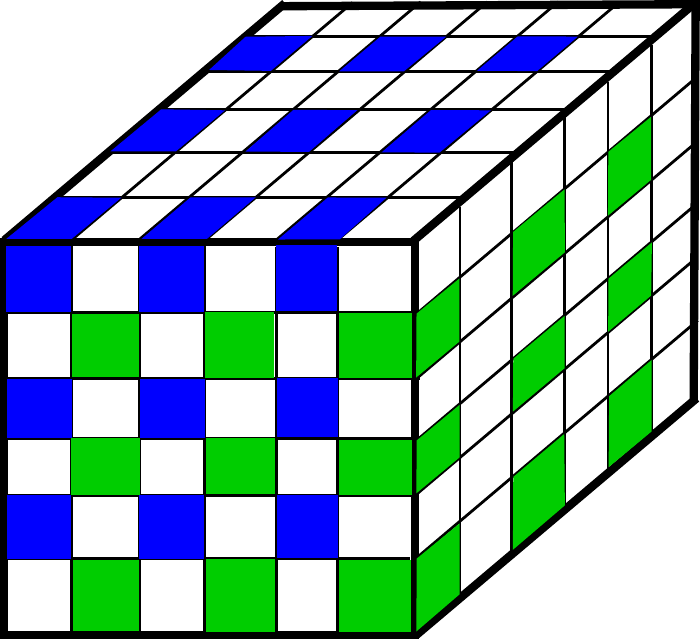}
  \caption{Provided the subdomains are larger than twice the
    bandwidth, the points contained in each blue boxes can be
    performed simultaneously.}
  \label{fig:PB-SYM-PD-str}
\end{figure}

As long as two points are separated by $2 h_s$ in space
dimension or by $2 h_t$ in time, the cylinders
induced by the two points will not overlap. PB-SYM-PD decomposes
the points in $A \times B \times C$ subdomain of identical
size. As long as these subdomains are larger than $2 h_s$ in a spatial
dimension and $2 h_t$ in the temporal dimension, the points
in domain $(x, y, z)$ can be done concurently with the points in
$(x + 2, y, z)$. See Figure~\ref{fig:PB-SYM-PD-str} for a graphical depiction.

We implemented the first version of this algorithm that organizes the
subdomain in 8 sets where the first set is composed of the subdomains
$(2i, 2j, 2k)$, the second of all the subdomains $(2 i + 1, 2 j, 2 k)$,
the third of all the subdomains $(2 i, 2 j + 1, 2 k)$, etc.. The
algorithm processes the sets the one after the other by doing each
subdomain within a set in parallel. Two such sets are depicted in
Figure~\ref{fig:PB-SYM-PD-str}. In practice, this is implemented using
$8$ OpenMP parallel-for constructs.\looseness=-1

\begin{algorithm}
\caption{\textsc{PB-SYM-PD}}
\begin{algorithmic}
\For{all points $i=(x_i,y_i,t_i)$}
  \State $localpoints[\floor{\frac{A X}{G_x}}][\floor{\frac{B Y}{G_y}}][\floor{\frac{C T}{G_t}}].add(x_i,y_i,t_i)$
\EndFor
\For{$abase \in \{0;1\}$}
  \For{$bbase \in \{0;1\}$}
    \For{$cbase \in \{0;1\}$}
      \State Do the work of the next three loops in parallel
      \For{$a = abase; a \leq A; a+=2$}
        \For{$b = bbase; b \leq B; b+=2$}
          \For{$c = cbase; c \leq C; c+=2$}
            \State Process points in $localpoints[a][b][c]$ using PB-SYM
          \EndFor
        \EndFor
      \EndFor
    \EndFor
  \EndFor
\EndFor
\end{algorithmic}
\label{alg:pb-sym-pd}
\end{algorithm}

\subsection{Coloring and Scheduling}

Note that such an implementation overconstrains the execution of the
algorithm. Indeed, despite they are not in the same set, subdomain $(1,
0, 0)$ and $(64, 64, 64)$ can be computed at the same time. The real
constraint is that the points contained in a subdomain can be
processed safely as long as none of the point contained in a
neighboring subdomain is being processed. Using the OpenMP tasking
construct with dependency introduced in OpenMP 4.0~\cite{OpenMP4}, one
can precisely express this constraint, which is a 27-point stencil
constraint.

Formally speaking, this problem is a combination of multiple very
classical graphs and scheduling problem. Modeling the subdomains as a
27-point stencil graph, identifying sets of subdomains to perform in
parallel is coloring the vertices of the graph so that no two
neighboring vertices share the same color. It differs from classical
graph coloring problems (surveyed in~\cite{GMP05}) in that the objective
is not to minimize the number of colors or to balance the number of
vertices of each color but to minimize the execution time of the
implied schedule.

\begin{figure}
  \centering
  \includegraphics[width=.5\linewidth]{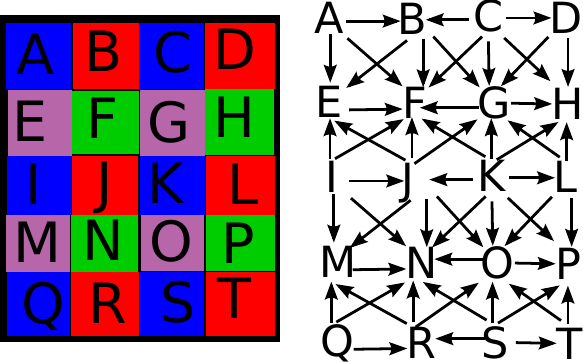}
  \caption{Dependency structure implied by a coloring of the
    subdomain. (Shown for a 2D decomposition for simplicity.)}
  \label{fig:PB-SYM-PD-graph}
\end{figure}

The implied schedule is actually a scheduling of a graph of dependency
which is extracted from the coloring. Each edge of the stencil graph
is oriented from the vertex of the lowest color to the vertex with the
highest color. Each vertex is associated with a processing time 
proportional to the number of points inside the sub-domain
the vertex represents and the neighboring subdomains. This
construction is represented in Figure~\ref{fig:PB-SYM-PD-graph}.

The OpenMP scheduler is not explicit in which order tasks that are
available are performed. Though it is reasonable to assume that the
tasks will be scheduled using a greedy algorithm. Therefore, the
classic guarantee from Graham's List Scheduling~\cite{Graham69} is
valid:
$$T_P \leq \frac{T_1 - T_\infty}{P} + T_{\infty},$$
where $T_1$ is the total amount of work to perform and $T_\infty$ is
the length of the longest chain in the dependency graph.

We propose to reduce the critical path by using a load-aware coloring
of the subdomains. We use a greedy coloring algorithm which 
considers the subdomains in a particular order and for each subdomain
will color it with the smallest color that does not conflict with the
already colored neighbors. Such greedy coloring algorithm are commonly
used in parallel computing applications~\cite{GMP05,Boman05-EuroPar,Deveci2016}. However, we will use
an ordering that color the vertices in non-increasing order of the
number of points contained in the subdomain. While this algorithm is a
heuristic, it is likely to obtain a coloring that minimizes the
critical path\footnote{The complexity of this problem is currently
  unknown to the authors. The general case of an arbitrary graph is
  NP-Hard by reduction to graph coloring. But the case of a 1D stencil
  is clearly polynomial.}. We call this algorithm to compute STKDE:
PB-SYM-PD-SCHED.

To decrease the critical path further, we propose an other algorithm
PB-SYM-PD-REP which replicates the subdomains that are on the critical
path (and its neighbors) of the graph so that the points of a
subdomain can be performed in parallel. As long as the critical path
is longer than $\frac{n}{2P}$, the tasks on the path are replicated an
additional time and the critical path is recomputed. Replication of
subdomain causes to have to initialize additional subdomains and to
reduce them which increases the work in a similar way
PB-SYM-DR. However, the replication should be limited to a few chains
in the graph and therefore should limit the overhead. Note that this
problem is similar to scheduling a DAG of moldable tasks~\cite{Lepere02,Hunold13}.

\section{Experiments}

\label{sec:experiments}

\subsection{Experimental Setting}

\paragraph{Execution environment} The machine used to perform the experiments is
composed of two Intel Xeon E5-2667 v3 processors for a total of
16 cores clocked at 3.2Ghz. Despite each core is capable of hosting 2
hyperthreads, hyperthreading was disabled. The machine is equipped with
128GB of DDR4 clocked at 2.133 Ghz. The node uses RHEL 7.2
and Linux 3.10. All codes are written in C++ and are compiled using GCC 5.3, which implements
OpenMP 4.0, with optimization flags \texttt{-O3 -march=native -fopenmp
  -DNDEBUG}. All reported execution times exclude all IOs.

\paragraph{Dataset} Four datasets are used in our experiments. Their
key attributes are summarized in Table~\ref{tab:instance}. 
Out first dataset, Dengue, is of epidemiological nature and consists of the space-time
locations of dengue fever cases reported to the “Sistema de Vigilancia
en Salud P\'{u}blica” for the city of Cali, Colombia during 2010 and 2011. The
system is updated on a daily basis with records of individuals that
have been diagnosed with dengue fever, including patient address and
the date of diagnosis. Addresses are standardized to a common address
format, and spelling and other syntactical errors are manually
corrected~\cite{delmelle2013spatio}; they are then geocoded and masked
to the nearest street intersection level to maintain
privacy~\cite{Kwan04}. In 2010, 9606 were successfully geocoded
(11,760 reported cases, or 81.7\%), whereas in 2011, 1562 cases were
successfully geocoded (1822 reported cases, or 85.7\%).

The PollenUS dataset is composed of 588K tweets of US users from
February 2016 to April 2016 related to Pollen (i.e., that mention
keywords such as ``pollen'' and ``allergy''). Tweets that did not
contain a precise localization were approximated by picking a random
location in the approximated region provided by
Gnip\footnote{\url{http://www.gnip.com/}}. Tweets are mostly located
in the contiguous continental US. Therefore, this is the region we
modeled with a spatial resolution ranging from $.2^\circ$ to $1^\circ$
and a temporal resolution of 1 day.

The Flu dataset was obtained from the Animal Surveillance
database of the Influenza Research
Database\footnote{\url{https://www.fludb.org/}}.
The dataset contains observations of birds tested positive for
carrying any strain the avian flu from 2001 to 2016. The birds were
observed all over the world and the modeled region spans as West as
Alaska and East as Japan, as South as the southern shore of Australia
and as North as the Northern shore of Russia. The domain is modeled
with spatial resolution from $.2^\circ$ to $1^\circ$ and a temporal
resolution of 1 day.\looseness=-1

The eBird\footnote{\url{http://ebird.org}} dataset is a collection of worldwide rare bird spotting
obtained from a crowdsourcing effort managed by Cornell's Lab or
Ornithology~\cite{ebird}. Despite the service was launched in 2002,
the database contains many older entries of birds that are known to
have been observed at the time. For our experiment, we use the last 20
years of bird spotting and model the entire world with a temporal
resolution of 3 days, and a spatial resolution of either $.1^\circ$ or
$.5^\circ$.

For all datasets, we created multiple instances of the problem coded
Lr, Mr, Hr for low, medium, and high resolution and Lb, Hb, VHb for
low, high and very high bandwidth. See details in Table~\ref{tab:instance}. 

\begin{table}
    \resizebox{\linewidth}{!}{\begin{tabular}{|l|rrr|rr|}
\hline
Instance & n & $G_x$ x $G_y$ x $G_t$ & Size & $H_s$ & $H_t$ \\
\hline
Dengue\_Lr-Lb & 11056 & 148x194x728 & 79MB & 3 & 1 \\
Dengue\_Lr-Hb & 11056 & 148x194x728 & 79MB & 25 & 1 \\
Dengue\_Hr-Lb & 11056 & 294x386x728 & 315MB & 2 & 1 \\
Dengue\_Hr-Hb & 11056 & 294x386x728 & 315MB & 50 & 1 \\
Dengue\_Hr-VHb & 11056 & 294x386x728 & 315MB & 50 & 14 \\
PollenUS\_Lr-Lb & 588189 & 131x61x84 & 2MB & 2 & 3 \\
PollenUS\_Hr-Lb & 588189 & 651x301x84 & 62MB & 10 & 3 \\
PollenUS\_Hr-Mb & 588189 & 651x301x84 & 62MB & 25 & 7 \\
PollenUS\_Hr-Hb & 588189 & 651x301x84 & 62MB & 50 & 14 \\
PollenUS\_VHr-Lb & 588189 & 6501x3001x84 & 6252MB & 100 & 3 \\
PollenUS\_VHr-VLb & 588189 & 6501x3001x84 & 6252MB & 50 & 3 \\
Flu\_Lr-Lb & 31478 & 117x308x851 & 117MB & 1 & 1 \\
Flu\_Lr-Hb & 31478 & 117x308x851 & 117MB & 2 & 3 \\
Flu\_Mr-Lb & 31478 & 233x615x1985 & 1085MB & 2 & 3 \\
Flu\_Mr-Hb & 31478 & 233x615x1985 & 1085MB & 4 & 7 \\
Flu\_Hr-Lb & 31478 & 581x1536x5951 & 20260MB & 5 & 7 \\
Flu\_Hr-Hb & 31478 & 581x1536x5951 & 20260MB & 10 & 21 \\
eBird\_Lr-Lb & 291990435 & 357x721x2435 & 2391MB & 2 & 3 \\
eBird\_Lr-Hb & 291990435 & 357x721x2435 & 2391MB & 6 & 5 \\
eBird\_Hr-Lb & 291990435 & 1781x3601x2435 & 59570MB & 10 & 3 \\
eBird\_Hr-Hb & 291990435 & 1781x3601x2435 & 59570MB & 30 & 5 \\
\hline
\end{tabular}
}
  \caption{Properties of the datasets. The domain is expressed in voxels
    and in MB. Bandwidth are expressed in voxels.}
  \label{tab:instance}
\end{table}

\subsection{Algorithm design}

\begin{table}
  \resizebox{\linewidth}{!}{\begin{tabular}{|l|rr|rrrr|r|}
\hline
& \multicolumn{6}{c|}{Time (in seconds)} & speedup\\
Instance & VB & VB-DEC & PB & PB-DISK & PB-BAR & PB-SYM & PB-SYM \\
\hline
Dengue\_Lr-Lb 	& 219.163 	& 2.283 	& 0.040 	& 0.029 	& 0.035 	& 0.028 	& 1.429 \\
Dengue\_Lr-Hb 	& 220.591 	& 13.878 	& 1.298 	& 0.564 	& 1.152 	& 0.499 	& 2.601 \\
Dengue\_Hr-Lb 	& 866.445 	& 9.522 	& 0.089 	& 0.082 	& 0.085 	& 0.084 	& 1.060 \\
Dengue\_Hr-Hb 	& 871.774 	& 55.206 	& 5.169 	& 2.272 	& 4.563 	& 2.074 	& 2.492 \\
Dengue\_Hr-VHb 	& 1056.172 	& 404.845 	& 51.885 	& 11.478 	& 42.994 	& 7.431 	& 6.982 \\
PollenUS\_Lr-Lb 	& 518.859 	& 7.639 	& 1.106 	& 0.347 	& 0.922 	& 0.256 	& 4.320 \\
PollenUS\_Hr-Lb 	& 12721.001 	& 189.337 	& 23.539 	& 7.700 	& 18.527 	& 4.708 	& 5.000 \\
PollenUS\_Hr-Mb 	& 17179.482 	& 3126.947 	& 357.743 	& 86.129 	& 295.791 	& 57.528 	& 6.219 \\
PollenUS\_Hr-Hb 	& 	& 	& 2666.104 	& 583.175 	& 2212.626 	& 382.566 	& 6.969 \\
PollenUS\_VHr-Lb 	& 	& 	& 2428.126 	& 1004.174 	& 1949.988 	& 759.722 	& 3.196 \\
PollenUS\_VHr-VLb 	& 	& 	& 603.789 	& 240.236 	& 488.388 	& 179.834 	& 3.357 \\
Flu\_Lr-Lb 	& 926.360 	& 3.691 	& 0.035 	& 0.032 	& 0.034 	& 0.032 	& 1.094 \\
Flu\_Lr-Hb 	& 966.328 	& 3.797 	& 0.081 	& 0.046 	& 0.070 	& 0.042 	& 1.929 \\
Flu\_Mr-Lb 	& 8591.165 	& 30.355 	& 0.305 	& 0.278 	& 0.298 	& 0.277 	& 1.101 \\
Flu\_Mr-Hb 	& 8957.175 	& 32.018 	& 0.714 	& 0.384 	& 0.608 	& 0.323 	& 2.211 \\
Flu\_Hr-Lb 	& 	& 536.091 	& 5.702 	& 5.089 	& 5.454 	& 5.059 	& 1.127 \\
Flu\_Hr-Hb 	& 	& 591.955 	& 12.795 	& 6.822 	& 10.992 	& 7.072 	& 1.809 \\
eBird\_Lr-Lb 	& 	& 	& 396.811 	& 147.951 	& 322.580 	& 125.248 	& 3.168 \\
eBird\_Lr-Hb 	& 	& 	& 6969.187 	& 1897.051 	& 5611.158 	& 1067.395 	& 6.529 \\
eBird\_Hr-Lb 	& 	& 	& 8373.273 	& 3226.016 	& 6470.764 	& 2229.460 	& 3.756 \\
eBird\_Hr-Hb 	& 	& 	& 	& 	& 	& 34577.745 	& \\
\hline
\end{tabular}
}
  \caption{Runtime of different algorithm. The speedup of PB-SYM over PB is given.  }
  \label{tab:seq_table}
\end{table}

Table~\ref{tab:seq_table} presents the runtime of the point-based
algorithms presented in Section~\ref{sec:algo-seq}. The runtime are
given in seconds. As expected for each dataset, the runtime increases
with resolution and bandwidth. Exploiting the disk invariant with
PB-DISK provides a large reduction in the runtime, even more so when
the temporal bandwidth is high. PB-BAR provides a more modest time
reduction.

\begin{figure}
  \centering
  \includegraphics[width=.8\linewidth]{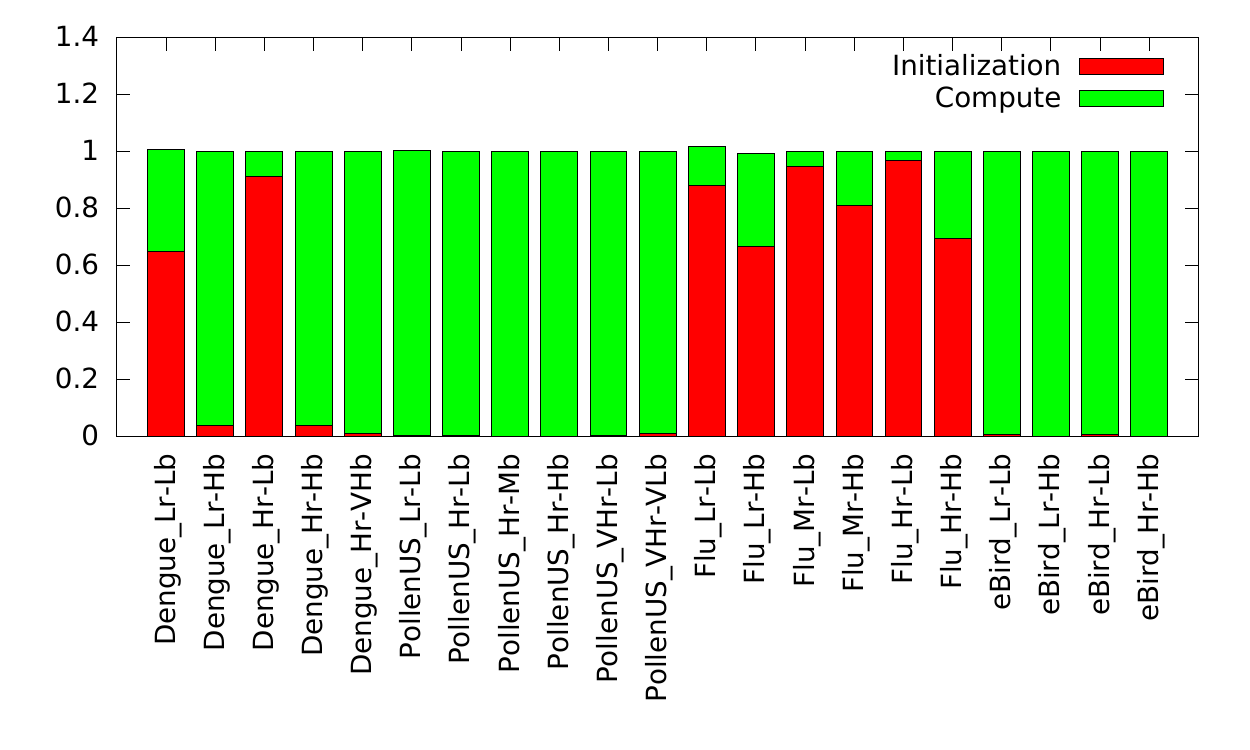}
  \caption{Breakdown of the runtime of PB-SYM. Some instances are
    mostly Initialization.}
  \label{fig:breakdown}
\end{figure}

PB-SYM combines the reductions of both PB-BAR and PB-DISK and achieves
the best performance in most cases. PB-SYM provides improvements up to
a factor of $6.969$ on the PollenUS\_Hr-Hb case. The cases where PB-SYM
provides little improvement are explained either by the instance having a
low bandwidth, and therefore there is little redundant computation to
take advantange of, or by the fact that the runtime is dominated by
initialization cost. The breakdown of the runtime of PB-SYM is given
in Figure~\ref{fig:breakdown} and shows that PB-SYM is composed of two
phases, initializing the memory to store the density estimation of the
voxels, and computing the density cylinder surrounding the
points. PB-SYM reduces the computation cost but initializes the memory
in the same way as PB. Initialization is a predominant runtime cost in
the Flu dataset since there were only 31K occurence of confirmed avian
flu recorded in the last 15 years but they span most of the earth.

Comparing to the gold implementation would not be fair because it is
in a different programming language. Though we implemented the VB
algorithm, and also a variant, VB-DEC, that partitions the points in
blocks of size equal to the bandwidth to only compute distances
between voxels and points that have a chance to have an impact on
it. The results are given in Table~\ref{tab:seq_table} and show that
even VB-DEC is usually at least an order of magnitude larger than PB
and typically two orders of magnitude larger than PB-SYM.

\subsection{Domain-based parallelism}

\paragraph{PB-SYM-DR}

\begin{figure}[tbh]
  \centering
  \includegraphics[width=1\linewidth]{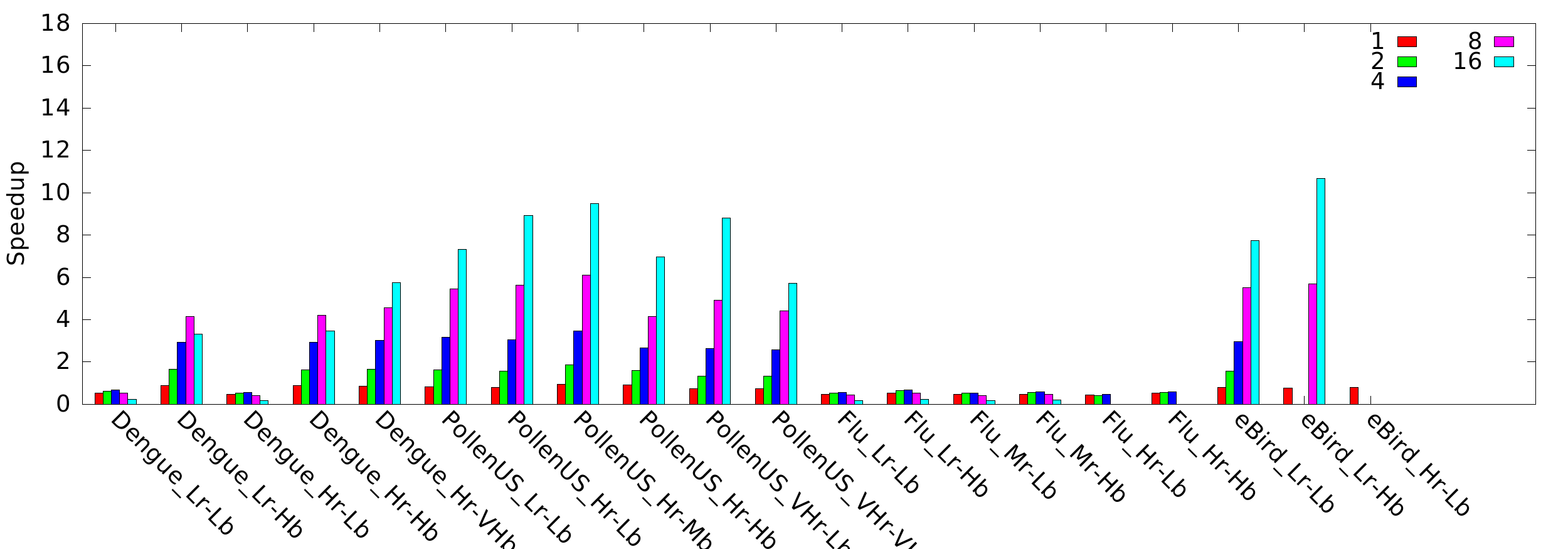}
  \caption{Speedup of PB-SYM-DR for different number of threads
    used. Some instances run out of memory.}
  \label{fig:sym-dr-speedup}
\end{figure}

The performance of the PB-SYM-DR algorithm is given in
Figure~\ref{fig:sym-dr-speedup}. Because it replicates the domain
space, PB-SYM-DR was not able to complete the calculations for
Flu\_Hr-Lb and Flu\_Hr-Hb when using 8 threads and 16 threads as the
memory requirement exceeds the 128GB available on the machine. None of
the high resolution eBird instances could have their domain
replicated.

All the instances that have a high initialization cost get a speedup
lesser than 1 since the threads spend their time allocating extra
memory and reducing the content of that extra memory. The only
instances to achieve a speedup higher than 8 when using 16 threads are 3 of 
the PollenUS instances and one low resolution eBird instances. Not
surprisingly, they are the instances with the largest fraction of
computation. (See Figure~\ref{fig:breakdown} for reference.)

\paragraph{PB-SYM-DD}

\begin{figure}[tbh]
  \centering
  \includegraphics[width=\linewidth]{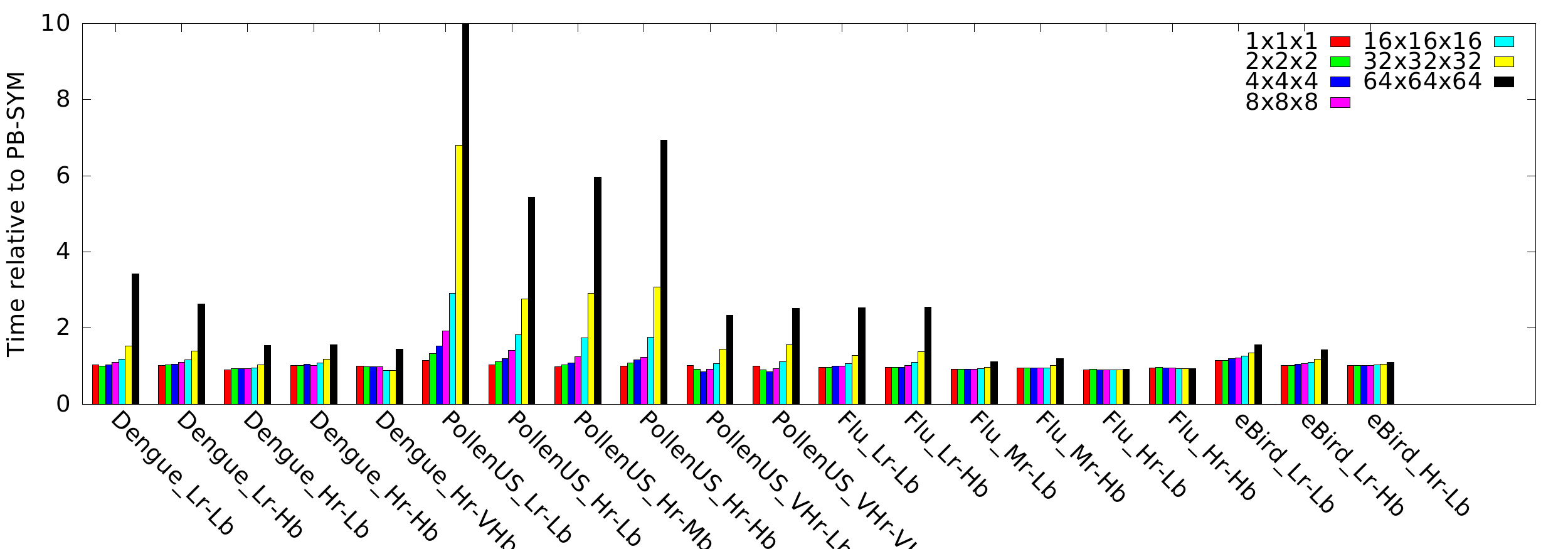}
  \caption{Overhead of PB-SYM-DD: runtime of PB-SYM-DD with a single
    thread normalized to PB-SYM. Some decomposition improve
    performance dues to better cache fitting. Overdecomposition can
    cause significant overhead.}
  \label{fig:pb-sym-dd-speedup}
\end{figure}

The PB-SYM-DD algorithm (described in Section~\ref{sec:pb-sym-dd}) also has some
overhead induced by cut cylinder surrounding a point. To measure the
overhead beforehand, we run the decomposition algorithm using
different number of subdomains ranging from a single subdomain (1x1x1)
to a very fine grain decomposition in 64x64x64 subdomains (except on the eBird\_Hr\_Hb where such a test is computationally expensive). The results
of that experiment is presented in
Figure~\ref{fig:pb-sym-dd-speedup}. The figure presents the overhead
of the decomposition relatively to PB-SYM. Two things are
apparent. Some decomposition can actually reduce the runtime by
providing better cache locality; for instance Flu\_Hr-Lb is 9.8\% faster
using a 16x16x16 decomposition than PB-SYM.

However, in most cases, the decomposition actually increases the
runtime. For instance, a 16x16x16 decomposition increases the amount
of work by 9\% in the Flu-Lr-Hb cases, and a 64x64x64 decomposition
induces an overhead of 254\%. Even on the Dengue\_Lr-Hb, a simple
4x4x4 decomposition already induces a 5\% overhead which grows to 69\%
for a 16x16x16 decomposition.  Such large decomposition are likely to
be necessary since the points in the datasets are often organized in
clusters: a decomposition in 16 subdomains is unlikely to provide a
load balanced execution. The PollenUS suffers from the highest overheads,
with a 16x16x16 decomposition leading to a 74\% overhead on PollenUS-Hr-Mb,
and a 495\% overhead for a 64x64x64 decomposition

\begin{figure}[tbh]
  \centering
  \includegraphics[width=\linewidth]{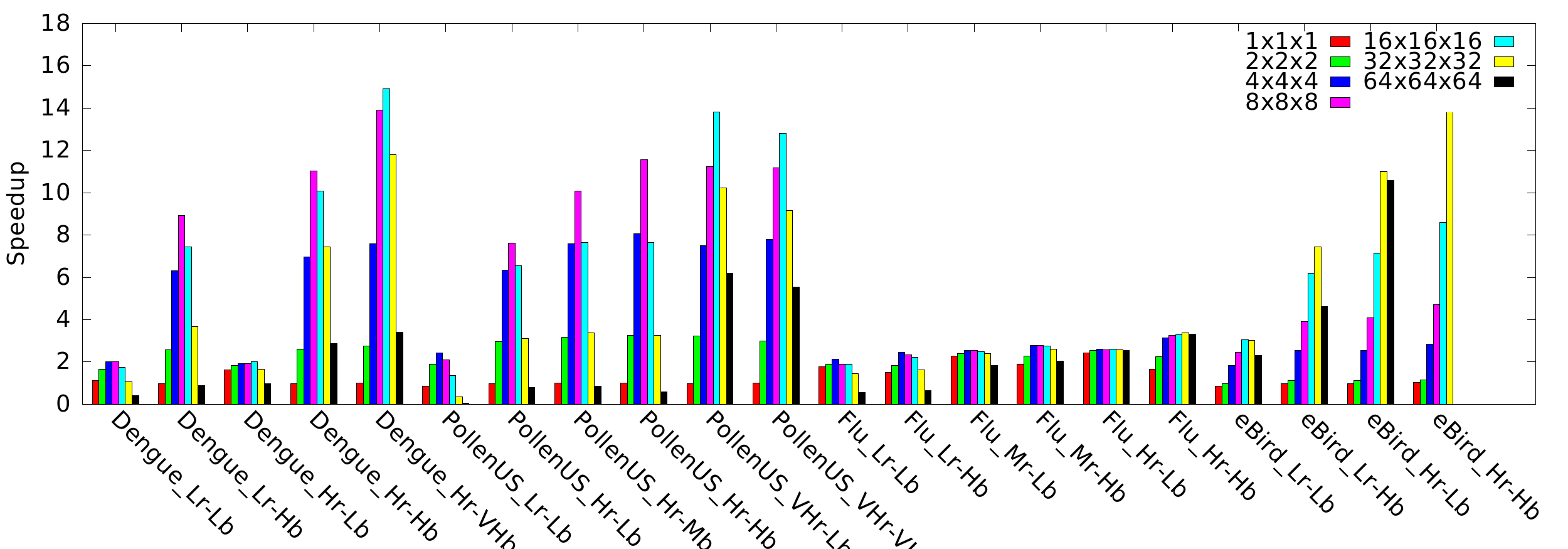}
  \caption{Speedup of PB-SYM-DD with 16 threads. While PB-SYM-DD achieves good
    speedup in some cases, the overdecomposition overhead usually
    hinders the performance.}
  \label{fig:pb-sym-dd-16-speedup}
\end{figure}

The speedup achieved by a parallel execution of PB-SYM-DD using 16
threads is presented in Figure~\ref{fig:pb-sym-dd-16-speedup}. This
algorithm achieves overall better speedup than PB-SYM-DR. It achieves a
speedup greater than 8 on 9 instances. Let us note that a speedup of
$14.9$ is achieved on Dengue\_Hr-VHb using a 16x16x16 decomposition
and of $14.8$ on eBird\_Hr-Hb with a 32x32x32 decomposition.

Despite a significant overhead on the PollenUS\_Hr instances, a speedup
of $10$ and $11.5$ can be observed on the Mb and Hb cases using 8x8x8
decomposition. This speedup is relatively impressive provided the
decomposition induces an overhead of 25\% and 23\% in these two
cases. The 8x8x8 decomposition in that case is still imbalanced and a
finer decomposition can reach perfect load balance; however the
overhead induced by the decomposition prevents taking advantages of
that better load balance. For PB-SYM-DD to be useful,
it is important to pick the decomposition to ensure load
balance without suffering from an unbearable computational overhead.

A more modest speedup between $2$ and $4$ can be observed on the
instances with higher initialization cost, such as on the Flu
dataset. Indeed, most of the time is spent on initializing the memory
to hold the density estimates. Even if the algorithm performs memory
initialization in parallel, there is not much room for parallelism in
memory operations. Even more so since the initialization touches the memory
for the first time, which makes the operating system
allocates the pages in physical memory. The speedup of the
initialization phase using 16 threads is about the same for all
instances and is at about $3$. The small speedup of PB-SYM-DD on
instances with high initialization cost is completely explained by
this phenomenon. Even if the compute phase was reduced to $0$, the
speedup on the entire calculation would only be $3.7$.

\subsection{Point-based parallelism}

\begin{figure}
  \centering
  \includegraphics[width=\linewidth]{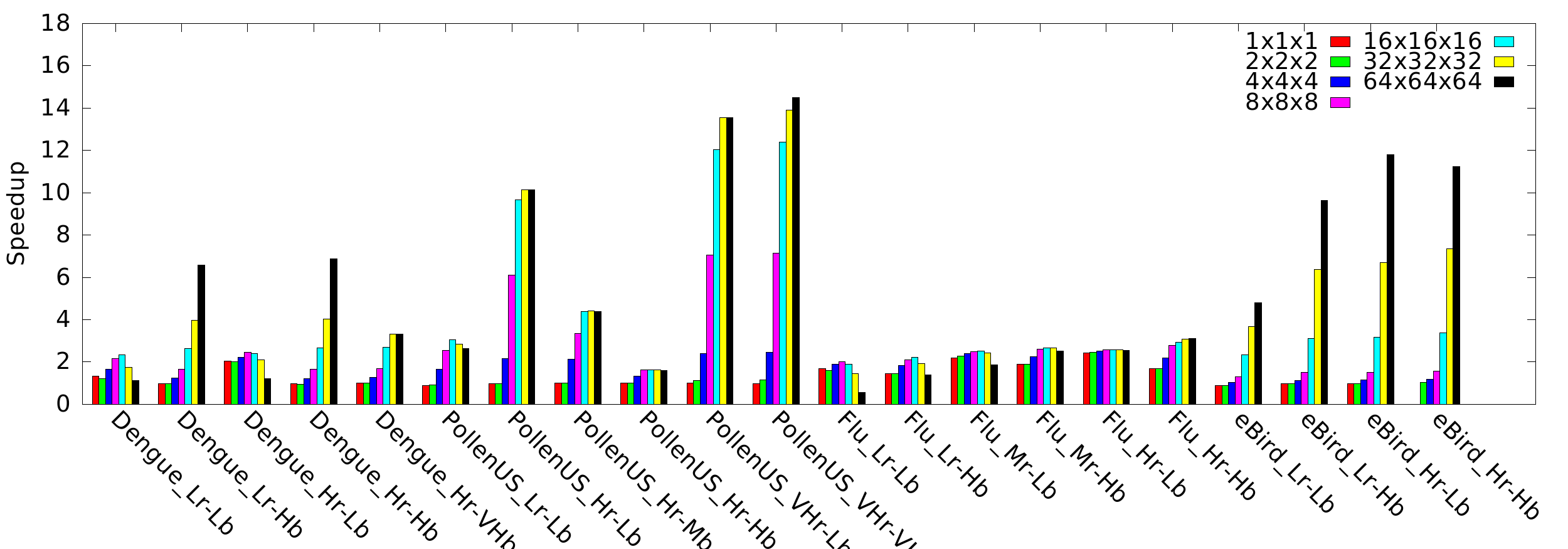}
  \caption{Speedup of PB-SYM-PD using 16 threads. Note that decompositions of
    subdomain smaller that twice the bandwidths are adjusted.}
  \label{fig:pb-sym-pd}
\end{figure}

The
performance of PB-SYM-PD is shown in Figure~\ref{fig:pb-sym-pd} for
different decomposition. Note that there is a requirement in PB-SYM-PD
that the subdomain be larger than twice the bandwidth (see details in
Section~\ref{sec:pb-sym-pd}). Therefore if the number of subdomain
leads to too small subdomains, the decomposition is adjusted to fit
the required size.

\begin{figure}
  \centering
  \includegraphics[width=\linewidth]{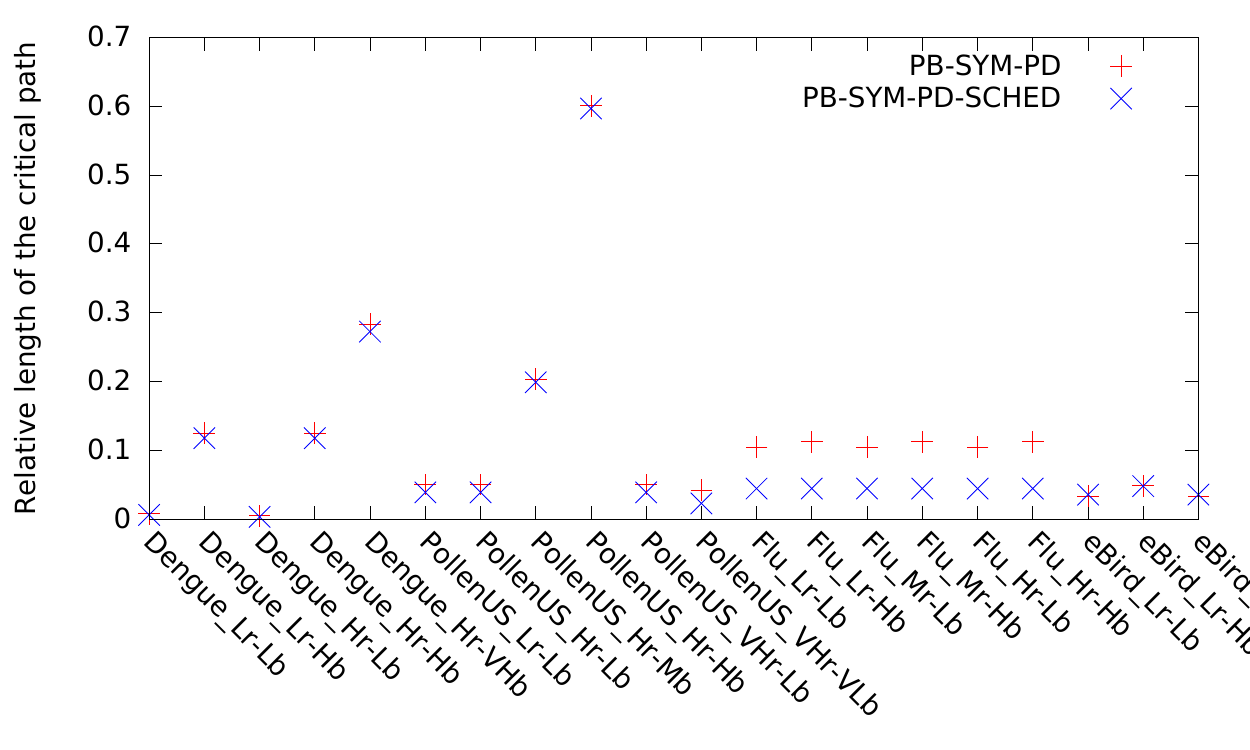}
  \caption{Length of the critical path of the parallel algorithms
    PB-SYM-PD and PB-SYM-PD-SCHED for a 64x64x64 decomposition}
  \label{fig:PB-SYM-PD-loadbalance}
\end{figure}

PB-SYM-PD does not scale well on all instances. The highest speedup achieved on
PollenUS\_Lr-Lb is $2.6$. It is interesting to notice
that the speedup usually increases with decomposition size. However,
the speedup is limited because of load
imbalance. Figure~\ref{fig:PB-SYM-PD-loadbalance} shows the implicit
critical path within the PB-SYM-PD algorithm for the highest
decomposition used in the experiment (64x64x64). Most instances have a
critical path of about 10\% of the total work. If Graham's bound is
reached, that implies a speedup lesser than $6.15$. 
PollenUS\_Hr-Hb has a critical path of 55\% which implies a speedup
lesser than $1.6$. However, the speedup is typically lower because of
the limited scalability in the initialization phase.

\begin{figure}
  \centering
  \includegraphics[width=\linewidth]{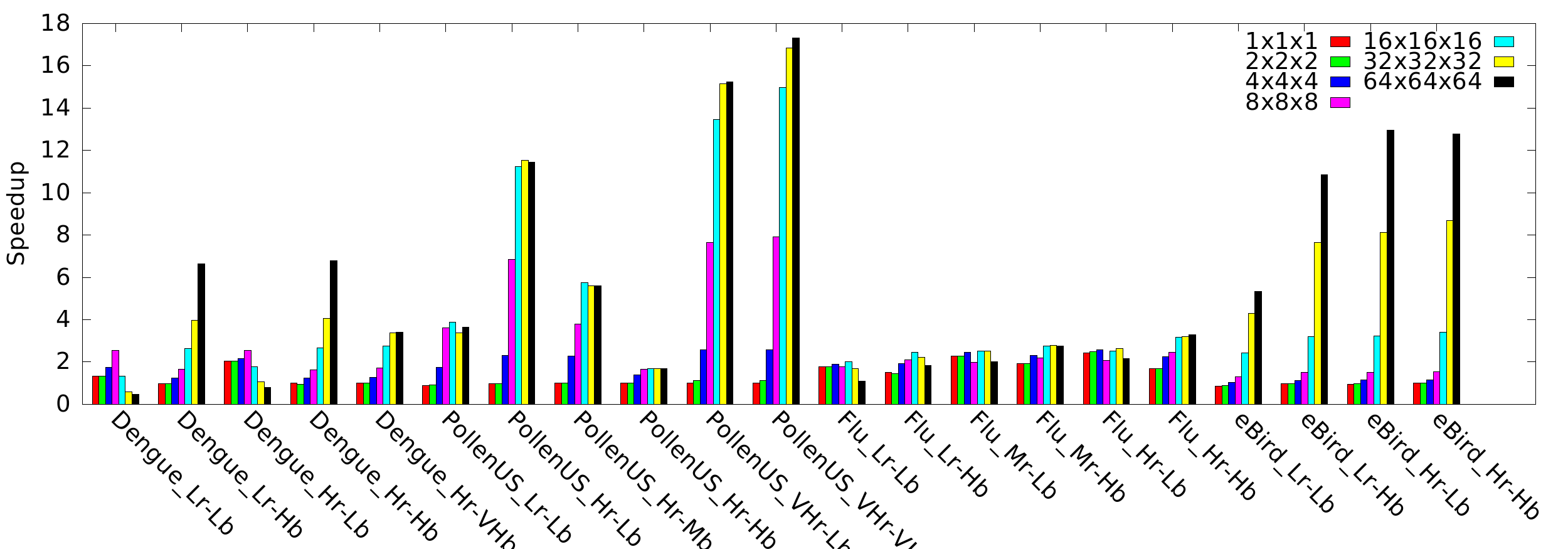}
  \caption{Speedup of PB-SYM-PD-SCHED with 16 threads.}
  \label{fig:PB-SYM-PD-SCHED-speedup}
\end{figure}

PB-SYM-PD-SCHED aims at improving the load balance by arranging the
execution to process the most loaded subdomain
first. Figure~\ref{fig:PB-SYM-PD-loadbalance} shows that
PB-SYM-PD-SCHED marginally decreases the critical path in all but one
case. Even if that decrease is marginal, PB-SYM-PD-SCHED ensures that
the highest loaded subdomain are executed first which tends to
construct a better execution in practice. And one can see from
Figure~\ref{fig:PB-SYM-PD-SCHED-speedup} that this scheduling improves
parallelism significantly for the PollenUS instance. A superlinear
speedup occurs on PollenUS\_VHr-VLb: it is due to the decomposition of
the points which improves the locality of the computation compared to
the sequential execution.

\begin{figure}
  \centering
  \includegraphics[width=\linewidth]{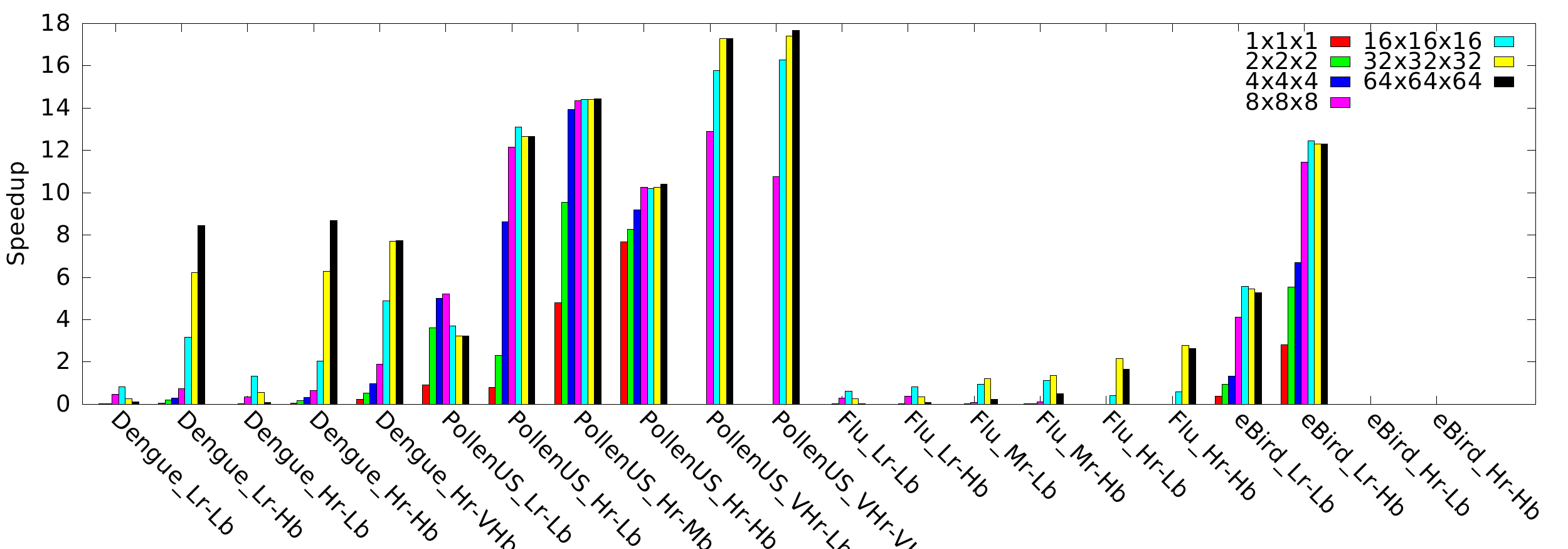}
  \caption{Speedup of PB-SYM-PD-REP using 16 threads for different
    decompositions. Flu\_Hr-Lb and Flu\_Hr-Hb run out of memory for
    small decomposition.}
  \label{fig:PB-SYM-PD-REP-speedup}
\end{figure}

However, the parallelism is still constrained by load
imbalance. PB-SYM-PD-REP performs domain decomposition on the
subdomain of the most loaded chains in the dependency graph until the
critical is small. Figure~\ref{fig:PB-SYM-PD-REP-speedup} shows the
speedup achieved by PB-SYM-PD-REP. Using PB-SYM-PD-REP achieves a
speedup larger than 8 on 8 instances. Though the speedup is close to 0
for small decompositions and on certain instances. Indeed, when there
is no (or little) decomposition, most of the domain is replicated and
the execution has similar drawbacks as PB-SYM-DR.

\subsection{Summary and Discussion}

\begin{figure*}
  \centering
  \includegraphics[width=\linewidth]{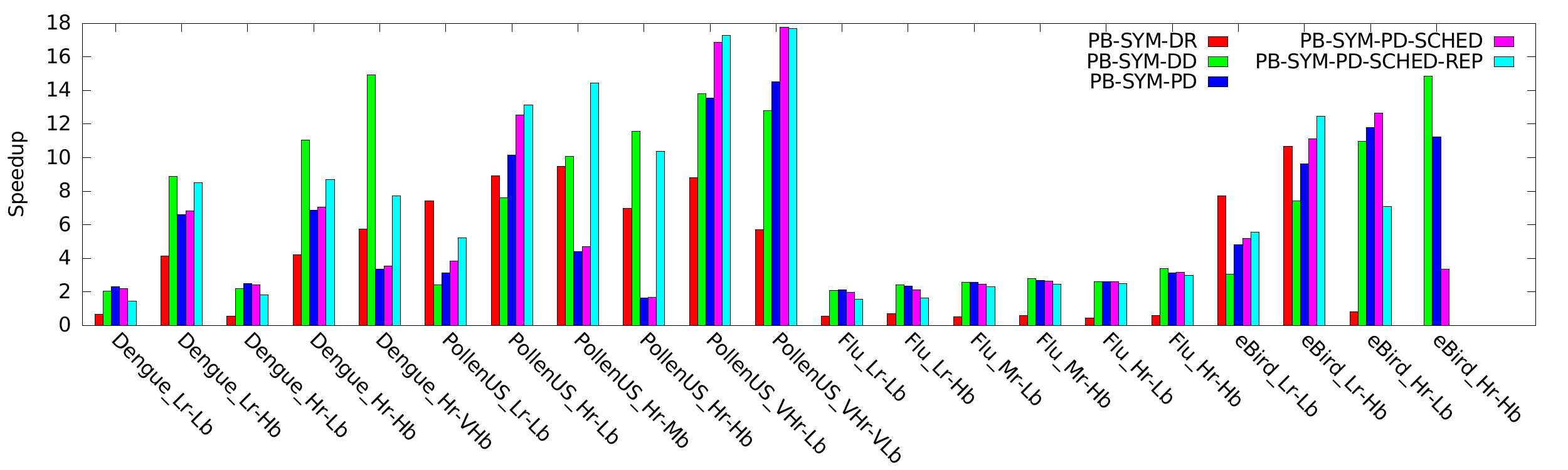}
  
  \caption{Best configurations}
  
  \label{fig:bestof}
\end{figure*}

Figure~\ref{fig:bestof} summarizes the performance of all our
methods. It shows the performance achieved by the best configuration
of each particular algorithm. On Dengue instances, PB-SYM-DD usually
leads to the best performance because of its low overhead on that
instance while maintaining a good load balance. On the PollenUS
instances, the smart scheduling of PB-SYM-PD-SCHED-REP is
necessary to reach the highest parallelism. The flu instances are
mostly dominated to initialization overhead because it is very sparse,
as such, PB-SYM-DR performs much worse than the other methods, and
past that issue, the method leads to little performance
difference. The eBird instances have a very small memory
initialization overhead because they are dense in computation. This
makes approaches that replicate the voxel space perform well at low
resolution, but runs out of memory at high resolution there.

What we need to do is to develop a parametric model for the problem
that will take into account memory availability, cost of memory
initialization, expected cost of computing the kernel density. Using
that model finding the best execution strategy becomes a combinatorial
problem.

\section{Related Works}

\label{sec:soa}

The PhD Thesis of U. Lopez-Novoa~\cite{LopezNovoa} is on the topic of
computing kernel density estimation for arbitrarily shaped
kernels. The crop and chop method presented revolves around cropping
the voxel space around a point with similar goal as PB, and compute
the density estimates in parallel on a multi-core CPU or on a GPU
before the data is aggregated back on the CPU. The arbitrarily shaped
density considered does not expose the invariants leveraged by PB-SYM
which makes their work inapplicable to STKDE. \cite{Eaglin17} considers
the same variant of kernel density that we consider and expand the
techniques of~\cite{LopezNovoa} by considering coalescing. However,
they still use a voxel based algorithm which is very slow (on a
resolution lower than our lowest resolution, the parallel computation
takes .6 seconds on a GPU which is slower than our reported sequential
time on a larger instance).

The two most common spatial problems solved with parallel computing
are N-body interactions~\cite{Aaseth03} and Particle in Cell
simulations~\cite{karimabadi06}. STKDE is different in the sense that there
is no iterating over the problem multiple times as we see in the
timesteps of N-body or particle in cell, no direct interaction between
points, and no field updates as in Particle in Cell (unless you
consider a field discretized at a one voxel size which makes the
analogy not quite useful).

The summation of kernel function, in particular of radial basis
function attracted some
attention~\cite{MarchXiaoYuBiros2015,FornbergLarssonFlyer2011,YangDuraiswamiGumerovDavis2003}. While
some method presented in these papers are similar to some of the
decomposition we discuss here, the mathematical property of the kernel
function are different. In particular the kernel functions that these
papers consider do not have the grid-aligned symmetries that the
space-time kernel density estimate have. That symmetry is what we
leverage in PB-SYM to gain an order of magnitude of acceleration and
that leads to a more complex management of parallelism we investigated.

Spatial applications can use spatial partitioning techniques 
such as recursive bisection~\cite{berger87}, jagged partition~\cite{ujaldon96,pinar97,DeveciRDC16}, or
rectilinear partition~\cite{nicol94,manne96}. Depending on the algorithms used for
STKDE, the objective function is different. A partitioning for
PB-SYM-DD needs a good load balance, to minimize the number of cut
cylinders. But PB-SYM-PD
needs a decomposition where the subdomains have a minimum size and the load balance critirion is harder to express since
neighboring subdomains can not be processed simultaneously.

\section{Conclusion}

We presented in this paper the space-time kernel density estimation
problem which is useful in the visualization of events located in
space and time. We proposed sequential algorithms to decrease the
complexity of the problem. And we investigated four parallel
algorithms, two of which are pleasingly parallel but at the cost of
not being work efficient. We then designed a work-efficient parallel
algorithm but the dependency structure tends to prevent high degree of
parallelism. Using graph coloring and moldable scheduling techniques
we made that latter parallel algorithm more parallel by adding some
work-overhead.

For each instance of the problem, one of the parallel algorithm
achieved an interesting performance. However, it is clear that we need
to model the instance and the platform to control the various overhead
and be able to pick the parallel strategy that will derive the highest
performance. It would also be interesting to look at distributed
memory machines and accelerators to reduce the runtime further since
real-time is desirable for interactive applications. In term of the
application, we would like to investigate how these methods apply to a
bandwidth that adapts to the density of population of the area is also
of interest.

\section*{Acknowledgment}

The authors would like to thank Dr. Daniel Janies for pointing us to
the Flu dataset.
Support from US NSF XSEDE Supercomputing Resource Allocation
(SES170007) "Accelerating and enhancing multi-scale spatiotemporally
explicit analysis and modeling of geospatial systems" is acknowledged.
This material is based upon work supported by the National Science Foundation under Grant No. 1652442.

\bibliographystyle{alpha}
\bibliography{report}

\end{document}